\documentstyle[11pt,paspconf,epsf]{article}
\def\edcomment#1{\iffalse\marginpar{\raggedright\sl#1\/}\else\relax\fi}
\reversemarginpar

\begin{document}
\title{Effects of NLTE and granulation on LiBeB abundance determinations }

\author{Dan Kiselman}
\affil{The Royal Swedish Academy of Sciences, Stockholm Observatory,
SE-133~60~~Saltsj{\"o}baden, Sweden}

\begin{abstract}
NLTE effects -- the errors caused by assuming local thermodynamic equilibrium (LTE) 
-- on LiBeB abundance determinations for cool stars are
discussed. These NLTE effects are significant in many cases for
B\,{\sc i} and Li\,{\sc i} lines. The Be\,{\sc ii} 313 nm lines are
not formed under LTE circumstances, but the NLTE effects on equivalent
widths seem to be rather small. Reasons for doing LTE abundance
analysis are discussed and confronted with the reasons for doing NLTE abundance analysis.
The impact of three-dimensional
photospheric inhomogeneity on line formation is discussed -- there are not many
definite results on this yet, but there may be soon.
\end{abstract}

\section{Introduction}
Elemental abundance ratios in stars are not observed -- they are
derived from observations. The standard procedure is to use
a model of the stellar photosphere, characterised by a few
fundamental parameters, and then model the radiative transfer in
the spectral line in question. One of the input parameters in
this is the elemental abundance which can be adjusted until
the synthetic spectrum (or just an equivalent width) fits
the observations.

A stellar photosphere is a complex and dramatic place, a
borderline between the hot inferno that is the inside of the
star and cold, dark space. Many approximations
have to be used when modelling the photosphere
and the spectral line formation processes. 
Typical abundance work uses one-dimensional LTE model photosphere
and LTE modelling of the spectral lines. This inevitably causes errors.
The question is how big these errors are.

I will discuss NLTE effects in the formation of LiBeB spectral lines
that are used for abundance
determinations. I will also give a brief discussion on
the possible effects of photospheric inhomogeneity.
Note that these discussions will not necessarily be valid for other stars
than those of solar type.

\section{NLTE line formation}
\subsection{General considerations}
The strength of a spectral line, from a given photosphere, is set by
the line source function and the line opacity.
Consider the formal solution to the equation of radiative transfer for
outgoing intensity along vertical rays:

$$ I_\nu = \int_0^\infty S_\nu(\tau_\nu) e^{-\tau_\nu} d\tau_\nu. $$

The source function $S_\nu$ has contributions from the spectral line
and from the continuum, and so has the opacity $\kappa_\nu$ which sets the optical
depth scale, $d\tau_\nu = \kappa_\nu dz$.
The line source function contains the ratio between the upper and
the lower level involved in the line transition. The line opacity is
proportional to the lower level population. In local thermodynamic
equilibrium -- LTE -- these quantities
are easily computed from the Boltzmann and Saha relations and are thus
functions only of the local kinetic temperature, the electron
pressure, and the elemental abundance. The line source function then equals
the Planck function. We know, however, that we cannot expect LTE to be valid in a
stellar photosphere. If we relax that condition, we must compute level
populations by solving the rate equations and
thus take into account collisional and radiative transitions among all
energy levels in all ionisation stages. That is, unless we have good
reasons to simplify the problem by cutting away levels. The radiative
rates depend on the radiation field in the atom's lines and continua
and thus couple different parts of the photosphere with different parts
of the atom and different parts of the spectrum. We thus have an
extensive problem to solve, which we call a non-LTE or NLTE problem.
The term NLTE can mean anything that is not LTE, but all NLTE work
discussed here has used the assumptions of statistical
equilibrium, Maxwellian particle distribution, and complete
redistribution of photons in frequency and angle.
A NLTE effect is then an error introduced by the LTE assumption or a
deviation from detailed balance in the transition rates.

Happily there are now efficient codes to
solve multi-level NLTE problems for trace elements in
stellar atmospheres. The big problem then becomes the availability
of atomic data. The situation is improving, partly thanks to the
Opacity and the Iron Projects (Seaton et al. 1994, Hummer et
al. 1993), but all data that one
would like to have is generally not readily available at the required
precision, especially for heavier atoms.

The solution of a NLTE problem consists of a very big set of data describing
the population numbers, the transition rates between the levels, and the
radiative field at various frequencies and atmospheric depths.
We have to devise ways to talk about departures from LTE and to describe
very complex situations in simple words. Remember that what is behind
these effects is just the ordinary excitation and ionisation processes
and not any strange new physics. Remember also that what interests us
is the difference between the NLTE and the LTE results. A small shift
in, for example, the ionisation balance can have large effects on
weak lines from the minority species.

For solar-type photospheres one can set up the following rules of
thumb:
\begin{itemize}
\item Towards the blue \& ultraviolet: $J_\nu > B_\nu$. For a line with a
source function behaving more or less like in the two-level atom
approximation ($S_\nu^l = (1-\varepsilon_\nu)J_\nu + \varepsilon
B_\nu$), this leads to an increased line source function:
{\em optical pumping} or {\em overexcitation}.
Levels with important continua in the blue may contribute to
{\em overionisation}.
\item Towards the red: $J_\nu<B_\nu$. This tends to lead to opposite
effects as those describe for the blue \& ultraviolet.
\item Strong lines suffer photon losses which pushes the line
source function down, in the red also for weaker lines. Such effects in neutral
species can lead to {\em photon suction} which forces the ionisation
balance downwards compared the Saha value.
\end{itemize}

Note that an increase in the source function means that the emerging
intensity increases. So increasing the line source function of an
absorption line will decrease the line strength.
Increasing the opacity by increasing the level population of
the line's lower level will of course increase the line
strength in a typical photosphere.

The NLTE effects on the spectra of Li, Be and B in solar-type stars
are of interest not only for the sake of abundance applications. 
These light atoms allow more or less
complete model atoms to be compiled, and they are (almost
always) true trace elements. Thus the complexity of the problem is
decreased and in modelling these spectral lines we ``only'' have to
worry about the elemental abundance, the photospheric structure
(decoupled from the NLTE problem under study) and
the quality of the atomic data. Sometimes the abundance is so low that
the problem simplifies even more because the lines and continua are so
weak. All this means that if we are ever to be
able to do perfect NLTE modelling of any spectral lines of solar-type stars
without resorting to 
tuning free parameters in the atomic or the photospheric models, LiBeB lines are
among the best candidates.


\subsection{Lithium}
The NLTE formation of Li\,{\sc i} lines in cool stars ($T_{\rm eff} \le
7500\,{\rm K}$) was investigated
by Carlsson et al. (1994) in an extensive grid of photospheric models.
Pavlenko et al (1995) and Pavlenko \& Magazz\`u (1996) have investigated
very cool stars.

Lithium is sometimes so abundant that the strong Li\,{\sc i} resonance lines show
the typical effects of line-source-function depression by photon
escape. This makes the line stronger than in LTE so that an LTE
analysis will overestimate the abundance. Also involved 
are overionisation effects and the counteracting photon-suction
effect. The resulting abundance corrections display a complicated
dependence on stellar parameters and lithium
abundance as can be seen in the illustrations of Carlsson et al. The
subordinate
lines show more moderate abundance corrections than the resonance
doublet and the corrections are sometimes of the opposite sign. 
This is an argument for checking lithium abundances also with these
lines, if possible.

\subsection{Beryllium}
The interesting lines for beryllium studies of cool stars are the
Be\,{\sc ii} resonance lines at 313 nm.

Beryllium has not been subject to as much NLTE work as lithium.
Garc{\'\i}a L\'opez et al (1995) and Kiselman \& Carlsson (1995) showed that
overionisation (similar in mechanism to that of B\,{\sc i} described
below) and overexcitation more or less cancel the NLTE effect
on the Be\,{\sc ii} 313 nm equivalent widths in solar-type stars. 
The line opacity increases from the overionisation but that is
countered by the increase of the line source function by optical
pumping. 
This means that while the line formation circumstances differ
significantly from LTE, the resulting equivalent width
does not. The situation prevails over some range of stellar parameters
because both effects are driven by approximately the same ultraviolet 
radiation field.

One may therefore say that it is
alright to assume LTE in analysing the Be\,{\sc ii} lines until we know more.
But note again that the lines are {\em not} formed in LTE!

\subsection{Boron}
I will discuss the NLTE line formation of B\,{\sc i} lines in some detail
since the effects investigated by Kiselman (1994) and Kiselman \&
Carlsson (1996) seem to be important for the boron abundances and their
interpretation for metal-poor stars. 

The resonance lines of the neutral boron atom are all in the
ultraviolet as is the case for beryllium. But a fundamental difference
to the beryllium case is of course that we are observing lines from
the neutral ionisation stage.
When the neutral boron abundance is low, as is mostly the case in
metal-poor solar-type stars, the NLTE processes affecting the line
that has hitherto been most observed -- the doublet at 250 nm -- are:
\begin{list}{--}{}
 \item pumping in the line itself that raises the line-source function and
 thus weakens the line.
 \item overionisation caused by the pumping in the ultraviolet resonance
 lines. (And to a lesser extent by $J_\nu>B_\nu$ in the photoionisation
 continua.) This also weakens the line.
\end{list}

Both these mechanisms tend to make the lines weaker than in LTE and
the LTE assumption will thus lead to an overestimation of the line strength or
an underestimation of derived boron abundances. This means that
NLTE corrections always are positive except for very boron-rich stars
where photon-loss effects in the resonance lines become important.

Note that these results were calculated using OSMARCS photospheric
models from Uppsala (Edvardsson et al. 1993)
and the NLTE corrections of Kiselman \& Carlsson (1996) 
should only be applied to LTE results from similar
models. If you do otherwise you are not being consistent, but are in
effect mixing different models. The end result may not be so bad in
practice since
e.g. the OSMARCS and the Kurucz (1993) models tend to give rather similar LTE
results in these studies (Garc\'{\i}a L\'opez et al. 1998), but one
should be aware that the introduction of this inconsistency is a
potential source of worry -- at least for purists.

Application of these results to boron observations has shown that
the NLTE effects are important for the interpretation of boron data of
metal-poor stars
(e.g. Edvardsson et al. 1994, Garc\'{\i}a L\'opez et al. 1998, Duncan et
al. 1997). The corrections increase in magnitude with increasing
effective temperature and decreasing metallicity and have the effect
of decreasing the slope in plots of stellar boron abundance as
function of metallicity.

Most of the atomic data used in the modelling 
can probably be regarded as rather sound. But
there could be surprises. Collisional ionisation is something that
definitely should be improved. Background opacities 
should be looked more upon -- this is highly relevant also for LTE
work. (For the Sun Mg\,{\sc i} and Fe dominates the background at
209~nm and Mg\,{\sc i} and H-bf at 249~nm. For very metal-poor solar
type stars various processes involving hydrogen dominates the
background at these wavelengths.)
Blending of the ultraviolet resonance lines with background lines
has a definite impact on the
departures from LTE. Now, when observations of the 209 nm doublet are coming, a
better linelist for this region should be included.
The 209 nm observations of Rebull et al. (1998) cause problems anyway
since they imply different abundances than those from 250 nm -- with or
without NLTE effects considered. More such input would be welcome!

These uncertainties notwithstanding it is clear that boron abundance
determinations of metal-poor solar-type stars using B\,{\sc i} lines
and assuming LTE will always be underestimates.

\section{LTE or NLTE?}

Can we trust the NLTE results discussed above? If we can, how come
most authors use LTE analyses? Why does anyone bother to publish NLTE
corrections to LTE results? Why aren't you all running NLTE codes?
I have tried to find the top ten reasons for doing LTE abundance analysis,
which are discussed below. They are followed 
by the corresponding list of reasons for doing NLTE analysis.

\subsection{Ten reasons to do LTE abundance analysis}

\begin{enumerate}
\item{ \em LTE is a robust approximation that has served us
     surprisingly well.} Indeed it has. But beware of selection
     effects. Cautious abundance analysts have learnt to avoid the most
     problematic lines and situations.
\item{ \em There are good LTE spectrum synthesis codes around.}
     Apparently NLTE codes are still more cumbersome to use and demand
     lots of input data.
\item{ \em It makes it easier to compare results with other authors.}
     This is an argument to publish LTE abundances for reference, but not
     to use them as final estimates.
\item{ \em It is inconsistent to model only the lines under analysis in NLTE.}
     Is consistency better than truth? Indeed we would want to model
     the whole spectrum in NLTE using 3D NLTE photospheric models. While we
     cannot do everything it is not wrong to do the improvements we
     can. Note that NLTE effects on abundance determinations
     can go in both ways, therefore there
     are no systematic effects to be expected a priori when comparing
     different lines and continua modelled under different assumptions.
\item{ \em Atomic data for NLTE work are unreliable or non-existing.}
     This was certainly true a number of years ago. But the argument
     is getting less and less valid, especially for lighter atoms.
\item{ \em NLTE results are sensitive to unknown or unreliable ultraviolet
     opacities.} Yes, but so are LTE results!
\item{ \em There are so many ways to make mistakes in NLTE work: this makes
     results unreliable.} The input data for a line in an LTE
     spectral line synthesis consists 
     of something like ten numbers (depending on damping treatment).
     The boron model atom used by Kiselman \& Carlsson (1996) 
     consists of no less than 6000 numbers.
     The results are not sensitive to most of these numbers -- but there is
     always a risk of a mistake or error. This means that all NLTE results
     must be carefully checked and that all opportunities for checking
     against different lines must be taken. But is it better to choose
     an approach that {\em may} give erroneous results before one that
     we {\em know} is wrong?
\item{ \em NLTE modelling is too complicated for me to undertake.}
Perhaps, but it is possible to have someone compute NLTE corrections.
\item{ \em It was good enough for my PhD supervisor, so it is good enough for me.}
\item{ \em I don't understand what this NLTE thing is about anyway.}
\end{enumerate}

\subsection{Ten reasons to do NLTE abundance analysis}

\begin{enumerate}
\item{ \em LTE is wrong.} And that is in fact all there is to it.
\item{ \em LTE is wrong.} 
\item{ \em LTE is wrong.} 
\item{ \em LTE is wrong.} 
\item{ \em LTE is wrong.} 
\item{ \em LTE is wrong.} 
\item{ \em LTE is wrong.}  
\item{ \em LTE is wrong.} 
\item{ \em LTE is wrong.} 
\item{ \em LTE is wrong.} Has anyone missed it now? Photospheric
radiation fields depart from Planckian and the radiative rates
are generally larger than collisional rates. So the atomic populations
are set non-locally by the radiation field and Saha-Boltzmann
equilibrium is unlikely. LTE {\em may} be a very good approximation
sometimes, but its validity must be checked with NLTE modelling except
for a few very special cases.
\end{enumerate}

\subsection{LTE abundances and NLTE abundances}
To me it is strange to discuss 'LTE abundances' and 'NLTE abundances'
separately as if they were quantities of different physical
significance. 
We are trying to estimate abundances from observed spectra
using as realistic models of the photosphere and the line-formation
processes as possible. The current status of NLTE computations is the
current status of understanding of the line formation. We know that
LTE is wrong in most cases.
It is not a question of choosing one or the other!

My recommendation is that you may still publish LTE LiBeB abundances because they
are useful when comparing data from different sources and because they
serve to illustrate the size of NLTE effects. But NLTE analysis or
NLTE corrected abundances should be used whenever possible 
for the abundances you use for astrophysics.

\section{ Granulation and similar photospheric inhomogeneity}

\begin{figure}
\plotone{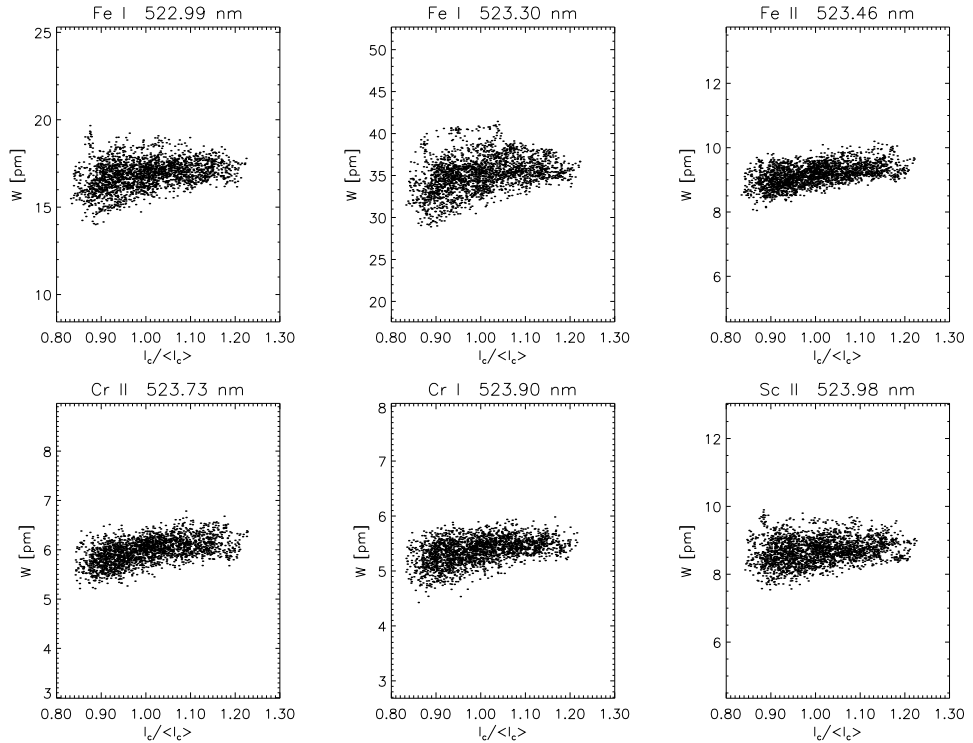} 
\caption{Equivalent widths of different spectral lines
as function of continuum intensity for
high-spatial-resolution solar observations. Note that all lines on
average grow
stronger over brighter regions, regardless of ionisation
stage.}
\end{figure}

Current model photosphere codes for cool stars (typically produced with the codes
of Kurucz or the Uppsala group) have a rather advanced treatment
of opacities taking thousands of wavelength points into consideration.
But they are still one-dimensional and plane-parallel and assume local
thermodynamic equilibrium (LTE). 

The solar photosphere is not plane-parallel and homogeneous like the
models. It shows (among other things) granulation -- a manifestation of
the convection that is driven by the strong cooling at the surface.

We know that other stars show similar surface structures. To what
extent does this affect abundance analysis?
If we look at how spectral lines behave over the solar granulation
pattern we notice that almost all (reasonable strong) 
lines grow stronger over the bright
granules with their strong temperature gradients and weaker in the
darker intergranular lanes (Holweger et al. 1990, Kiselman 1994).
Figure 1 exemplifies this for a number of spectral lines observed with
the Swedish Vacuum Solar Telescope on La Palma. Each point in these
plots corresponds to a single position in the solar granulation
pattern at disk centre. The spatial resolution is better than 1''.

This shows directly that the simple view
that dark and bright regions in the solar photosphere are like 1D
models with lower and higher effective temperatures is wrong.
The observation does not tell us directly how big
errors the use of 1D models will give, but it shows that line ratios, from
which abundance ratios are deduced, do not vary wildly -- an indication
that the effect is not too dramatic. There have been some attempts to
quantify errors of using standard 1D models by comparison with 3D or
2D simulations (e.g. Nordlund 1984, Bruls \&
Rutten 1992, Atroshchenko \& Gadun 1994, Kiselman \& Nordlund
1995, Gadun \& Pavlenko 1997). A problem is that the hydrodynamic
models necessarily have a simplified treatment of the radiative
transfer and so it is not evident if any differences are due to these
simplifications or to actual 2D/3D-effects.

Impressive advances have been made lately in three-dimensional hydrodynamic
modelling of the solar photosphere and one can definitely say that
solar granulation is
understood (e.g. Spruit 1997, Stein \& Nordlund 1998). 
So far only the Sun and a few similar stars
have been
modelled in this detail -- expect progress on this soon.

\subsection{Li\,{\sc i} lines in granulation}

One can imagine various ways for 1D models to be inadequate as
approximation of the real 3D structure. First the 1D temperature and
pressure structure may be bad representations of the spatial and
temporal mean of the 3D reality. This would then give errors also within
the LTE paradigm. Then there could be 3D NLTE effects such as the one
proposed by Kurucz (1995) for the Li\,{\sc i} resonance doublet formation
in extreme halo stars, which would cause severe errors in standard 1D
LTE abundance analysis. The idea was that hot ultraviolet
radiation from bright photospheric regions (something similar to 
granules) ionised the lithium
in the cool regions from which the Li\,{\sc i} lines would otherwise
have been strong. 

A problem with such a scenario is that the simple bifurcation of
the stellar photosphere into hot and cool regions is probably much
too simplified. Then the NLTE mechanisms are not so simple so that
one can just take for granted that more ultraviolet radiation leads to
significant overionisation. Finally, detailed modelling is needed to
find out how this works.

\begin{figure}
\plotone{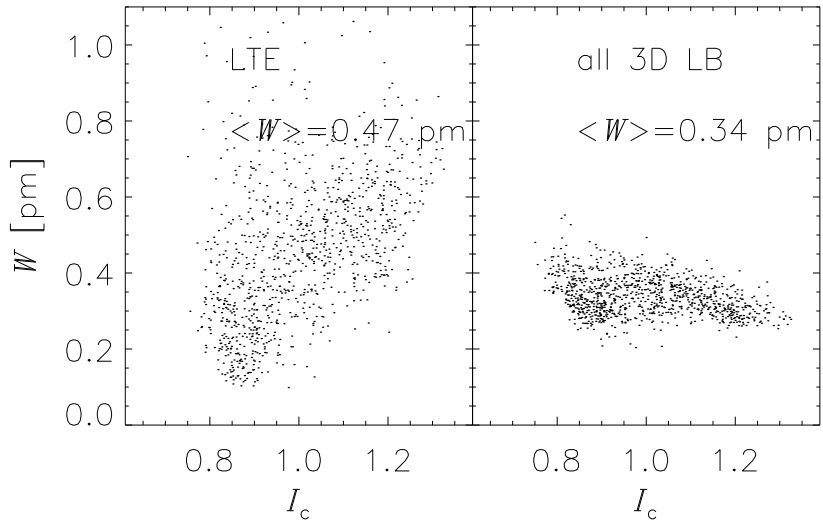} 
\caption{ Li\,{\sc i} equivalent widths as function of continuum
intensity in a solar granulation simulation snapshot. The left panel shows results
assuming LTE, right panel shows NLTE results with a (somewhat
simplified) treatment of 3D radiative transfer. Data from Kiselman (1997).}
\end{figure}

Uitenbroek (1998) and Kiselman (1997) investigated the NLTE Li\,{\sc
i} line
formation in 3D solar granulation simulation snapshots. Note that these
were models of the sun, not of metal-poor stars. The same type of model was used
but different methods were used for the line computations, and the interpretation
of the results is a little different in the two papers. But
it is clear that the 3D NLTE effects in this case are small. 
I must also point out that numerical experiments showed that the ultraviolet
radiation in the ionisation edges is not what determines the
ionisation balance. It is instead set by pumping in the bb-transitions
and thus at longer wavelengths where the radiation contrast is lower
than in the ultraviolet. This can be taken as an argument against 
the existence of very
large such 3D overionisation effects also in other stars than the Sun.

Martin Asplund in Copenhagen is currently working on 3D
hydrodynamic simulations of halo
star photospheres to investigate lines used for abundance analysis.
He finds a significant decrease of the derived lithium abundance for
these stars (Asplund 1998, private communication)
but warns that NLTE effects and other problems may be to blame for this.

According to the solar simulations displayed in Fig. 2, 
the LTE prediction is that the line gets
stronger over bright granules and that it shows a significant spread
for a given continuum intensity. This is not seen in the solar
observations of Kiselman (1998).
However, the NLTE prediction does not fit the observations
perfectly either. The observed $I-W$ relation is flat while the NLTE
simulation predicts a slight slope. I hesitate,
however, to draw any definite conclusions from this. The problem is that the line
is very weak and systematic effects in the continuum placement could
cause spurious slopes here. So I just end by proposing that
spatially resolved solar 
observations can be valuable in testing NLTE spectral-line modelling since we see
the lines of an element with one single abundance from different kinds of
photospheric structures. Note that in the Li\,{\sc i} case, the NLTE effect
in integrated light is rather small compared to the difference in slope and
spread in seen in Fig. 2. Perhaps it could be possible to
fine-tune the atomic models using these kinds of solar observations
and then use them on other stars where NLTE effects might be greater.

\section{Conclusions}

Correct your lithium and boron abundances for NLTE effects!

A more thorough NLTE investigation of beryllium would be welcome, until then LTE
is permissible and probably not too bad.

Results from extensive, detailed, and realistic modelling of the
impact of photospheric 3D 
structures on line formation for a variety of stars is soon
coming. That will be interesting. These studies must, however, be
investigated from the NLTE line formation point of view.

My guess is that effects of photospheric inhomogeneity will for
abundance analysts be like the 1D NLTE effects.
That is, they will not often be very dramatic but almost always of
significance as the demand on accuracy in abundance determinations grows.
But for some lines and situations the effects will
be large -- these cases must be looked for.

\acknowledgments

I thank Martin Asplund for sharing preliminary results and Luc Rouppe
van der Voort for comments on the manuscript.

\end{document}